\documentclass{article}
\usepackage{graphicx} % Required for inserting images
\usepackage{mathtools}
\usepackage{amsmath}
\usepackage{amssymb}
\usepackage{authblk}
\DeclareMathOperator\arctanh{arctanh}

\begin{document}

\title{Analysis of Drop Collisions}
\author{Alexis Tzelilis\\{\small Supervised by: Peter Lewin-Jones and Prof. James Sprittles}}

\date{
Department of Mathematics, University of Warwick %Dirty Trick :P
\\
\today
}
\maketitle

\begin{figure}[h]
    \centering
    \includegraphics[width=0.5\linewidth]{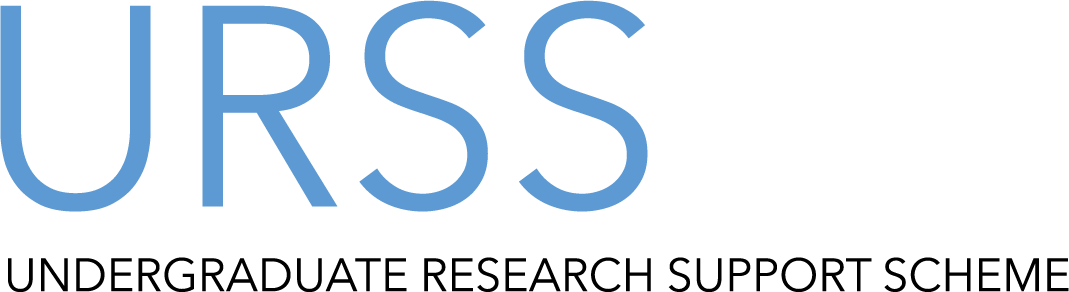}
\end{figure}

\tableofcontents

\section{Introduction}
When do droplets merge and when do they bounce? Over the last 10 years, advances in experimental techniques, such as high-speed cameras, have enabled us to make important discoveries on how the dynamics of thin gas films can influence the behaviour of liquid droplets. However, the exact physical mechanisms governing these phenomena are a subject of debate, leading to a variety of theoretical, experimental and computational approaches. In this document, I will study two problems: the head-on collision of two droplets, and the impact of a drop on a solid surface. I will outline the conventional fluid mechanical description, how it fails, and various candidates for the ``missing physics''.

\section{Conventional Fluid Mechanical Description}

\subsection{Lubrication Theory}
Right before collision, there is a thin layer of gas between the two drops, or between the drop and the solid. This layer of gas plays a big role in the behaviour of the drops. Therefore, we need to study the flow in the gas, and we do this using lubrication theory. Lubrication theory is used to study the behaviour of fluids where the characteristic length scale is significantly smaller in one dimension than the others (in this case, the height of the gas film).

Let $H$ be the characteristic vertical length scale and $L$ the characteristic horizontal length scale in the direction of the fluid flow. In the lubrication approximation we assume that the layer of fluid is very shallow such that

\begin{equation}
    \epsilon = \frac{H}{L} \ll 1
\end{equation}
\begin{figure}[h]
    \centering
    \includegraphics[width=0.5\linewidth]{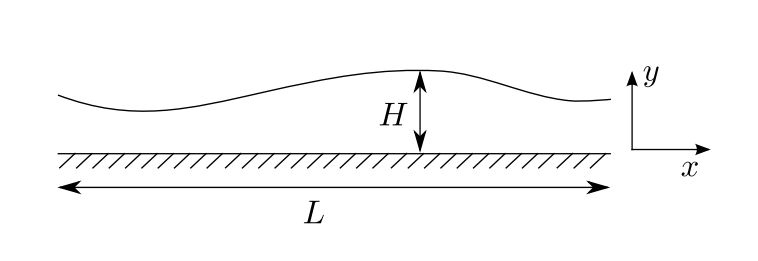}
    \caption{Sketch of flow in thin layer}
\end{figure}

In addition, let $U$ be the the scale of the horizontal velocity $u$. From incompressibility ($\mathbf{\nabla} \cdot \mathbf{u} =0$), the characteristic scale of the vertical velocity $v$ is $UH/L$.

\subsubsection{Dimensional Analysis}
I will now derive the lubrication equations in the gas from the Navier-Stokes equations by nondimensionalisation. The Navier-Stokes equations (ignoring gravity) are
\begin{equation}
    \frac{D\mathbf{u}}{Dt} \vcentcolon= \frac{d\mathbf{u}}{dt}+(\mathbf{u}\cdot\nabla)\mathbf{u} = -\nabla p +\mu \nabla^2\mathbf{u}
\end{equation}
where $\mathbf{u}(\mathbf{r}, t)$ is the velocity of the fluid at position $\mathbf{r}$ and time t, $p(\mathbf{r},t)$ is the pressure and $\mu$ the viscosity of the fluid. We now introduce non-dimensional variables
\begin{equation*}
    t=\frac{L}{U}\tilde{t}, \quad x=L\tilde{x}, \quad y=H\tilde{y}, \quad u=U\tilde{u}, \quad v=\frac{UH}{L}\tilde{v}, \quad p=P\tilde{p}
\end{equation*}
In 2D, the Navier-Stokes equations for the x-component becomes
\begin{equation*}
    \frac{\rho U^2}{L}\frac{D\tilde{u}}{D\tilde{t}} = -\frac{P}{L}\frac{\partial \tilde{p}}{\partial \tilde{x}} + \frac{\mu U}{L^2}\frac{\partial^2\tilde{u}}{\partial\tilde{x}^2} + \frac{\mu U}{H^2}\frac{\partial^2\tilde{u}}{\partial\tilde{y}^2}
\end{equation*}
Dividing by $\mu U/H^2$ the equation becomes
\begin{equation*}
    \frac{UH^2}{\nu L}\frac{D\tilde{u}}{D\tilde{t}} = -\frac{PH^2}{\mu UL}\frac{\partial\tilde{p}}{\partial\tilde{x}} + \frac{H^2}{L^2}\frac{\partial^2\tilde{u}}{\partial\tilde{x}^2} + \frac{\partial^2\tilde{u}}{\partial\tilde{y}^2}
\end{equation*}
Where $\nu = \mu/\rho$ is the dynamic viscosity. Therefore, we can see that inertia can be neglected if
\begin{equation*}
    Re\frac{H^2}{L^2} = \frac{UL}{\nu}\frac{H^2}{L^2} \ll 1
\end{equation*}
We expect the pressure to be significant so choosing pressure scale $P=\mu UL/H^2$.
Now
\begin{equation*}
    \epsilon^2Re\frac{D\tilde{u}}{D\tilde{t}} = -\frac{\partial\tilde{p}}{\partial\tilde{x}} + \epsilon^2\frac{\partial^2\tilde{u}}{\partial\tilde{x}^2} + \frac{\partial^2\tilde{u}}{\partial\tilde{y}^2}
\end{equation*}
Where $\epsilon = H/L$. So to first order in $\epsilon$ the horizontal Navier-Stokes equation can be approximated as
\begin{equation*}
    \frac{\partial\tilde{p}}{\partial\tilde{x}} = \frac{\partial^2\tilde{u}}{\partial\tilde{y}^2}
\end{equation*}
The Navier-Stokes equation for the y-component is
\begin{equation*}
    \frac{\rho U^2 H}{L^2}\frac{D\tilde{v}}{D\tilde{t}} = -\frac{\mu UL}{H^3}\frac{\partial \tilde{p}}{\partial \tilde{y}} + \frac{\mu UH}{L^3}\frac{\partial^2\tilde{v}}{\partial\tilde{x}^2} + \frac{\mu U}{HL}\frac{\partial^2\tilde{v}}{\partial\tilde{y}^2}
\end{equation*}
Multiplying both sides by $H^3/\mu UL$ we get
\begin{equation*}
    \frac{UH}{\nu}\frac{H^3}{L^3}\frac{D\tilde{v}}{D\tilde{t}} =-\frac{\partial \tilde{p}}{\partial \tilde{y}} + \frac{H^4}{L^4}\frac{\partial^2\tilde{v}}{\partial\tilde{x}^2} + \frac{H^2}{L^2}\frac{\partial^2\tilde{v}}{\partial\tilde{y}^2}
\end{equation*}
giving
\begin{equation*}
    \epsilon^4Re\frac{D\tilde{v}}{D\tilde{t}} =-\frac{\partial \tilde{p}}{\partial \tilde{y}} + \epsilon^4\frac{\partial^2\tilde{v}}{\partial\tilde{x}^2} + \epsilon^2\frac{\partial^2\tilde{v}}{\partial\tilde{y}^2}
\end{equation*}
Therefore, up to leading order, (setting $\epsilon = H/L=0$), we find that the pressure in independent of y. Thus, back in dimensional variables, we end up with the following lubrication equations.
\begin{align*}
    \frac{\partial p}{\partial x} &= \mu \frac{\partial^2 u}{\partial y^2}
    \\
    \frac{\partial p}{\partial y} &= 0
\end{align*}

\subsection{Forces Exerted by Gas Films During Drop Collisions}

\subsubsection{Flow of Gas Film}
\textbf{Drop-Solid Collisions.} Due to the radial symmetry of the problem, I will be working in cylindrical coordinates. Consider a spherical drop of radius $a$ approaching a wall at velocity $-\dot{d}\mathbf{z}$, where $d(t)$ is the smallest distance between the sphere and the wall. The height function is given by $h(r,t)$.
\begin{figure}[h]
    \centering
    \includegraphics[width=0.5\linewidth]{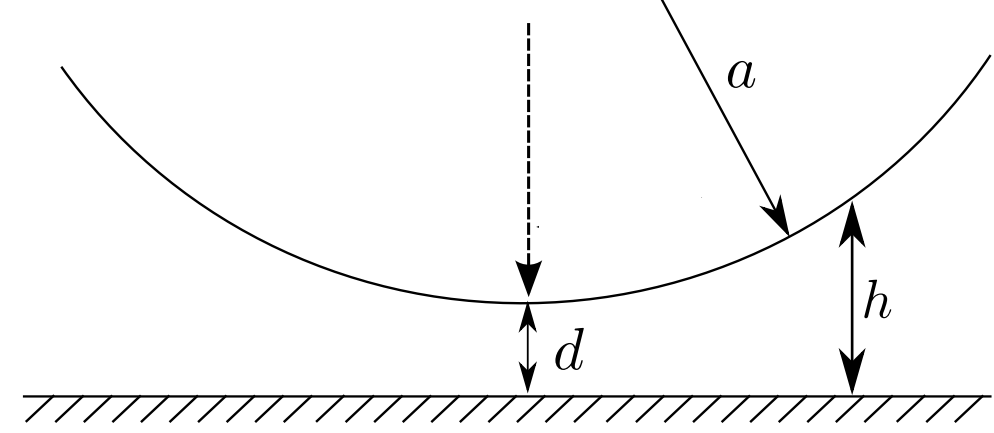}
    \caption{Diagram of Spherical Drop Approaching Wall}
\end{figure}

Firstly, assume that the flow is only radial, $(u_r,u_\phi,u_z)=(u,0,0)$. Using the equations from lubrication theory, in cylindrical coordinates, we get
\begin{equation*}
    \frac{\partial p}{\partial z} = 0
\end{equation*}
which means that $p$ and $\partial_r p$ are independent of z. Additionally, due the rotational symmetry of the problem, $p$ and $\partial_r p$ are independent of $\theta$. Note that the lubrication equation in cylindrical coordinates is
\begin{equation} \label{flow_equation}
    \frac{\partial p}{\partial r} = \mu \frac{\partial^2 u}{\partial z^2} \Rightarrow u=\frac{1}{2\mu} \frac{\partial p}{\partial r}z^2 + Az + B
\end{equation}
Using the no-slip boundary conditions $u(z=0)=0$ and $u(z=h) = U_r$, where $U_r(r,t)$ is the radial speed of the liquid at the gas-liquid interface, we get
\begin{equation*} 
    u = \frac{U_r z}{h} + \frac{1}{2\mu} \frac{\partial p}{\partial r}z(z-h)
\end{equation*}
\textbf{Drop-Drop Collisions.}
Consider two spherical drops, both of radius $a$, approaching the $z=0$ plane at velocity $-\frac{1}{2}\dot{d}\mathbf{z}$ and $\frac{1}{2}\dot{d}\mathbf{z}$ respectively. As before, $d(t)$ is the minimum distance between them and $h(r,t)$ is the distance function.
As before we can solve for u and using the no-slip boundary conditions $u(z= \pm h/2)=U_r$ we get
\begin{equation*}
    u=U_r + \frac{1}{2\mu}\frac{\partial p}{\partial r}(z^2 - \frac{1}{4} h^2)
\end{equation*}

\subsubsection{Mass-Flow Rate of Gas}
\textbf{Drop-Solid Collisions.} The mass-flow rate is given by
\begin{equation} \label{Q_drop_solid}
    Q = \int_{0}^{h} u \,dz\ = -\frac{h^3}{12\mu} \frac{\partial p}{\partial r} + hU_r/2
\end{equation}
\textbf{Drop-Drop Collisions.} The mass-flow rate is given by 
\begin{equation} \label{Q_drop_drop}
    Q = \int_{-h/2}^{h/2} u \,dz\ = -\frac{h^3}{12\mu} \frac{\partial p}{\partial r} + hU_r
\end{equation}

\subsubsection{Gas Forces}
\textbf{Mass Conservation.} From conservation of mass, we get
\begin{equation}\label{mass_conservation}
    \frac{\partial h}{\partial t} + \frac{1}{r}\frac{\partial (rQ)}{\partial r} = 0 \Rightarrow \frac{\partial (rQ)}{\partial r} = -\dot{d}(t)r \Rightarrow rQ=-\frac{1}{2}\dot{d}r^2
\end{equation}
Substituting $Q$ as calculated in (\ref{Q_drop_solid}) and (\ref{Q_drop_drop}), assuming $U_r=0$, we get
\begin{equation*}
    -\frac{rh^3}{12\mu}\frac{\partial p}{\partial r} = -\frac{1}{2}\dot{d}r^2 \Rightarrow \frac{\partial p}{\partial r} =\frac{6\mu\dot{d}r}{h^3}
\end{equation*}
Integrating with respect to $r$ we get
\begin{equation} \label{pressure_integral}
    p = \int \frac{6\mu\dot{d}r}{h^3} \,dr\
\end{equation}
\textbf{Geometry.} As the droplets approach the wall/each other, they start deforming.
\begin{figure}[h]
    \centering
    \includegraphics[width=0.5\linewidth]{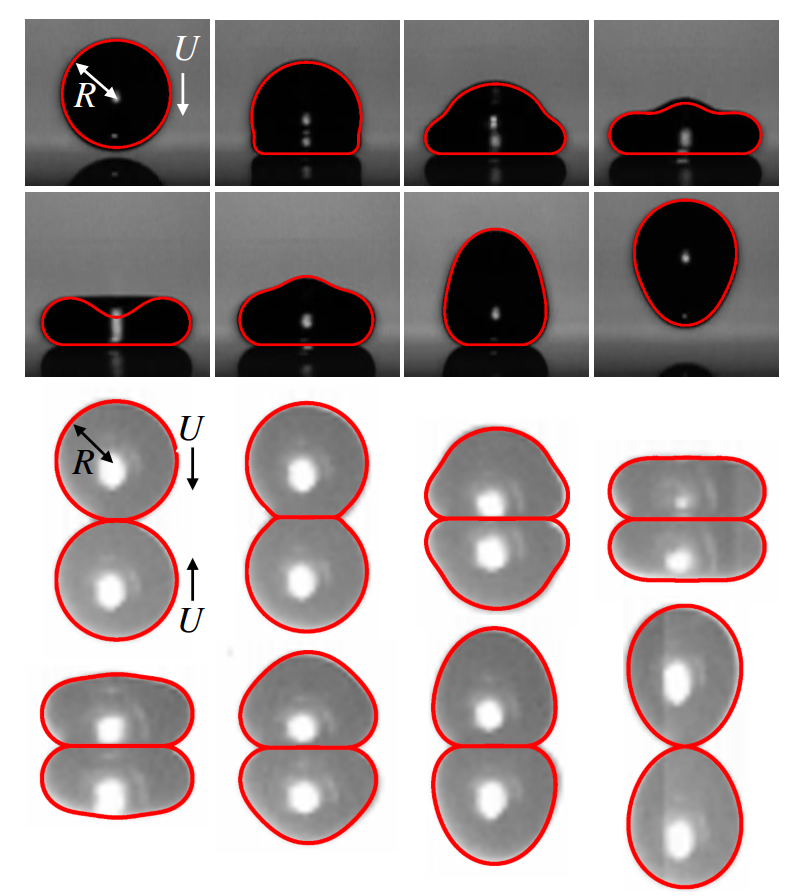}
    \caption{Upper: drop-solid bouncing of a R = 0.69 mm silicone oil droplet at Wel = 2.2, from de Ruiter
et al. (2015) \cite{deRuiter}. Lower: drop-drop bouncing of a R = 0.17 mm tetradecane droplet at Wel = 4.7
from Pan et al. (2008) \cite{Pan}. Red outlines are Peter Lewin-Jones' numerical calculations.}
\label{drop_experiments}
\end{figure}
As you can see from Figure \ref{drop_experiments}, the droplets flatten during impact, changing the geometry of the height function considered in equation (\ref{pressure_integral}). In our analysis, we will consider discs with height function
\begin{equation}\label{disc_height}
    h(r, t) = d(t)\quad(r,\theta)\in[0,a]\times[0,2\pi]
\end{equation}
We will also consider spheres with height functions
\begin{equation*}
    h(r, t) = d(t) + n(a - \sqrt{a^2 - r^2})\quad(r,\theta)\in[0,a]\times[0,2\pi]
\end{equation*}
where $n\in \{1,2\}$ is the number of droplets in the problem. However, to simplify our calculations, we will use the first two terms of the Taylor expansion of $\sqrt{a^2-r^2} \approx a - r^2/2a$ to get
\begin{equation}\label{sphere_height}
    h(r, t) = d(t) + nr^2/2a
\end{equation}
\textbf{Forces on Spheres.} Substituting (\ref{sphere_height}) into (\ref{pressure_integral}) and using $p(r\rightarrow\infty) = p_0$ we get
\begin{equation*}
    p = \int \frac{6\mu\dot{d}r}{(d+nr^2/2a)^3} \,dr\ = -\frac{3\mu\dot{d}a}{n(d+nr^2/2a)^2} +p_0
\end{equation*}
Finally, we integrate to get the force
\begin{equation} \label{spheres_force}
    F = \iint p-p_0 \,dA\ = -\frac{6\pi\mu\dot{d}a}{n} \int_{0}^{\infty}\frac{r}{(d+nr^2/2a)^2} \,dr\ =-\frac{6\pi\mu\dot{d}a^2}{dn^2}
\end{equation}
\textbf{Forces on Discs.} Substituting (\ref{disc_height}) into (\ref{pressure_integral}) and using $p(r=a)=p_0$ we get
\begin{equation*}
    p = \int \frac{6\mu\dot{d}r}{d^3} \,dr\ = p_0 + \frac{3\mu\dot{d}(r^2-a^2)}{d^3}
\end{equation*}
Integrating over the surface we get
\begin{equation} \label{discs_force}
    F = \iint_{Disc}p-p_0\,dA\ =\frac{6\pi\mu\dot{d}}{d^3} \int_{0}^{a}r(r^2-a^2)\,dr\ = -\frac{3\pi\mu\dot{d}a^4}{2d^3}
\end{equation}

\subsubsection{Model Breakdown}
From now on, let us consider only the drop-drop case. We can use Newton's Second Law to construct an ODE for the equation of motion of the two droplets. Note that $\Ddot{d}$ is the relative acceleration of the spheres, meaning that the acceleration of each sphere is $\frac{1}{2}\Ddot{d}$. Also note that in our calculations, $d(t)\geq0$ and $\dot{d}(t)\leq 0 \ \forall t\geq0$ \\
\textbf{Spheres} Substituting (\ref{spheres_force}) into Newton's 2nd Law we get
\begin{alignat*}{2}
    &&\frac{1}{2}m\Ddot{d}
    &= -\frac{3\pi\mu\dot{d}a^2}{2d}
    \\
    &\Rightarrow
    &\Ddot{d}
    &= -\frac{3\pi\mu a^2}{m} \frac{d (\ln{d})}{dt}
    \\
    &\Rightarrow
    &\dot{d}
    &= -\frac{3\pi\mu a^2}{m} \ln{d} + C
    \\
    &\Rightarrow
    &\dot{d}(t)
    &= \dot{d}(0) - \frac{3\pi\mu a^2}{m} \ln{\left(\frac{d(t)}{d(0)}\right)}
\end{alignat*}
It is possible to show that $d(t)>0 \ \forall t \geq 0$. Firstly, note that $\dot{d}(t)=0$ if and only if
\begin{alignat*}{2}
    && \dot{d}(0)
    &= \frac{3\pi\mu a^2}{m} \ln{\left(\frac{d(t)}{d(0)}\right)}
    \\
    &\iff
    &d(t)
    &= d(0)\exp{(\frac{\dot{d}(0)}{k})} > 0
\end{alignat*}
This means that the droplets come to a stop before they get close enough to merge (i.e. at $d=0$). But in reality, impacting drops do merge. Clearly, something is missing from the model described so far.\\
\textbf{Discs} The situation is worse when we take deformation into account. Substituting (\ref{discs_force}) into Newton's 2nd Law we get
\begin{alignat*}{2}
    &&\frac{1}{2}m\Ddot{d}
    &=-\frac{3\pi\mu\dot{d}a^4}{2d^3}
    \\
    &\Rightarrow
    &\Ddot{d}(t)
    &= \frac{3\pi\mu a^4}{2m}\frac{d}{dt}(\frac{1}{d(t)^2})
    \\
    &\Rightarrow
    &\dot{d}(t)
    &=\frac{3\pi\mu a^4}{2m}\frac{1}{d(t)^2} + C
    \\
    &\Rightarrow
    &\dot{d}(t)
    &=\dot{d}(0) + \frac{3\pi\mu a^4}{2m} (\frac{1}{d(t)^2} - \frac{1}{d(0)^2})
\end{alignat*}
Again it is possible to show that $d(t)>0 \ \forall t \geq 0$.
\begin{alignat*}{2}
    &&\dot{d}(t)
    &=0
    \\
    &\iff
    &\dot{d}(0)
    &=\frac{3\pi\mu a^4}{2m} (\frac{1}{d(0)^2} - \frac{1}{d(t)^2})
    \\
    &\iff
    &\frac{1}{d(t)^2}
    &=\frac{1}{d(0)^2}-\frac{2m\dot{d}(0)}{3\pi\mu a^4}
    \\
    &\iff
    &d(t)
    &=(\frac{1}{d(0)^2}-\frac{2m\dot{d}(0)}{3\pi\mu a^4})^{-1/2}>0
\end{alignat*}
Again, we can see that the droplets come to a stop before contact is made.

\section{Additional Physics}
The conventional framework fails. Contact is prevented, as also confirmed by simulations with deformable interfaces (see Pan et al. (2008) \cite{Pan} for the drop-drop case and Kolinski et al. (2014) \cite{Kolinski} for the drop-solid case).
This section considers various candidates for the ‘missing physics’.

\subsection{Gas-Kinetic Effects}
The conventional framework assumes that the mean free path, $\lambda$, of the gas molecules is much smaller than all relevant length scales in the problem. However, at atmospheric pressure, $\lambda \approx$ 70nm, so once the film height reaches 100s of nanometers thick, a common occurrence, we expect gas-kinetic effects to become important (see J Sprittles (2024) \cite{AnnualReview}).

\subsubsection{Navier-Slip Boundary Condition}
We can model the gas-kinetic effects by using the Navier-slip boundary conditions at $z=\pm h/2$ for the drop-drop scenario:
\begin{equation*}
    l\frac{\partial u}{\partial z}  = u
\end{equation*}
where $l$ is the slip length (See E Langa et al. (2007) \cite{NavierSlip}), which for gases is approximately equal to $\lambda$.  Plugging these into (\ref{flow_equation}) we get
\begin{equation*}
    u = \frac{1}{2}\frac{\partial p}{\partial r}(z^2-lh-\frac{h^2}{4})
\end{equation*}
We integrate to obtain the mass-flow rate
\begin{equation*}
    Q = \int_{-h/2}^{h/2} u \,dz\ = -\frac{h^2(h+6l)}{12\mu} \frac{\partial p}{\partial r}
\end{equation*}
Substituting this into (\ref{mass_conservation}) we get
\begin{equation*}
    \frac{h^2(h+6l)}{12\mu} \frac{\partial p}{\partial r} = \frac{1}{2}\dot{d}r \Rightarrow \frac{\partial p}{\partial r} = \frac{6\mu\dot{d}r}{h^2(h+6l)}
\end{equation*}
\textbf{Forces on Spheres.} Substituting from (\ref{sphere_height}) and integrating we get
\begin{align}\label{gke_pressure}
    p &= 6\mu\dot{d}\int\frac{r}{(d+r^2/a)^2(d+6l+r^2/a)} \,dr\ \nonumber
    \\
    &= \frac{\mu\dot{d}a}{6l^2}\left(\arctanh{\left( \frac{3l}{d+3l+r^2/a} \right)}-\frac{3l}{d+r^2/a}\right) + p_0
\end{align}
Note that we used $p\rightarrow p_0$ as $r\rightarrow\infty$ as the boundary condition. Integrating over the surface of the sphere we get the force exerted by the gas
\begin{align}\label{gke_force}
    F &= \iint p-p_0 \,dA\ \nonumber
    \\
    &=\frac{\mu\dot{d}a\pi}{3l^2}\int_{0}^{a} r(\arctanh{\left( \frac{3l}{d+3l+r^2/a} \right)} - \frac{3l}{d+r^2/a}) \,dr\ \nonumber
    \\
    &=\frac{\mu\dot{d}a\pi}{6l^2}\left[3l(\log{\left( 1+\frac{6l}{d+a} \right)}-\log{\left(1+\frac{6l}{d}\right))}\right. \nonumber
    \\
    &\quad\left. +(d+a)\arctanh{\left(\frac{3l}{d+3l+a}\right)}-d\arctanh{(\frac{3l}{d+3l})}\right]
\end{align}
In the limit $l\rightarrow0$ we recover the equations we got in the last chapter. Note that, using Taylor expansion around $l=0$ we get
\begin{equation}\label{arctanh_taylor}
    \arctanh\left(\frac{3l}{x+3l}\right) = \frac{3l}{x} - \frac{9l^2}{x^2} + \frac{36l^3}{x^3} + O(l^4)
\end{equation}
\begin{equation}\label{log_taylor}
    \log\left(1+\frac{6l}{x}\right) = \frac{6l}{x} - \frac{18l^2}{x^2} + O(l^3)
\end{equation}
for all $x>0$. Using $x=d+r^2/a$ in (\ref{arctanh_taylor}) and substituting into the pressure equation we get.
\begin{align*}
     p &= p_0 + \frac{\mu\dot{d}a}{6l^2}\left(\frac{3l}{d+r^2/a}-\frac{9l^2}{(d+r^2/a)^2} + \frac{36l^3}{(d+r^2/a)^3} + O(l^4)\right)
     \\
     &= p_0 - \frac{3\mu\dot{d}a}{2(d+r^2/a)^2} + \frac{6\mu\dot{d}a}{(d+r^2/a)^3}l + O(l^2)
     \\
     &=p_{(no-slip)} + \frac{6\mu\dot{d}a}{(d+r^2/a)^3}l + O(l^2)
\end{align*}
Similarly, we can use the appropriate $x$ values and substitute into the force equation to get
\begin{align*}
    F &= \frac{\mu\dot{d}a\pi}{6l^2}\left[3l\left(\left(\frac{6l}{d+a}-\frac{18l^2}{(d+a)^2}+O(l^3)\right)-\left(\frac{6l}{d}-\frac{18l^2}{d^2}+O(l^3) \right) \right)\right.
    \\
    &\quad+(d+a)\left(\frac{3l}{d+a}-\frac{9l^2}{(d+a)^2}+\frac{36l^3}{(d+a)^3}+O(l^4) \right)
    \\
    &\left.\quad-d\left(\frac{3l}{d}-\frac{9l^2}{d^2}+\frac{36l^3}{d^3}+O(l^4)\right)\right]
    \\
    &=\frac{\mu\dot{d}a\pi}{6}\left(-\frac{9a}{d(d+a)}+\frac{18(2ad+a^2)}{d^2(d+a)^2}l\right) + O(l^2)
\end{align*}
Since $d\ll a$ we have $d+a\approx a$ and $2ad+a^2 \approx a^2$
\begin{align*}
    \therefore F &\approx -\frac{3\mu\dot{d}a\pi}{2d} + \frac{3\mu\dot{d}a\pi}{d^2}l + O(l^2)
    \\
    &=F_{(no-slip)} + \frac{3\mu\dot{d}a\pi}{d^2}l + O(l^2)
\end{align*}
Therefore, we see $O(l)$ convergence to the no-slip equations as $l\rightarrow0$.
As you can see, the inclusion of gas kinetic effects introduces an attractive force, which could amend the problem of non-coallescence. We can repeat similar calculations to obtain the force on deformed droplets.

\subsection{Van der Waals Forces}
Films on solids or free liquid films are destabilized at sufficiently small scales by van der Waals (vdW) forces, that can drive interfaces into contact. Within a lubrication framework vdW forces can appear as a disjoining pressure
\begin{equation*}
    p_d = -\frac{A}{6\pi h^3}
\end{equation*}
which acts on each interface (see J Sprittles (2024) \cite{AnnualReview}). Here $A$ is the Hamaker constant for the particular liquid/solid–gas–liquid system considered, typically of the order of $10^{-21}–10^{-18}$J.

\textbf{Forces on Spheres} Integrating around the bottom surface of the sphere we get
\begin{align*}
    F_{vdW} &= \iint_{Drop}-\frac{A}{6\pi h^3} dA
    \\
    &= -\frac{A}{3}\int^R_0\frac{r}{(h_0+r^2/R)^3}dr
    \\
    &= -\frac{A}{3}\left[-\frac{R}{4(h+r^2/R)^2} \right]^R_{r=0}
    \\
    &= -\frac{AR}{12}\left(\frac{1}{h^2}-\frac{1}{(h+R)^2} \right)
\end{align*}
We can repeat similar calculations to obtain the forces on discs.

\section{Final Remarks}
In conclusion, we showed that the conventional lubrication model is insufficient for describing both head-on collisions of two droplets and the impact between a droplet and a solid surface. Additions to the model, such as the inclusion of Gas-Kinetic Effects and the van der Waals force, are required to fully explain these phenomena, in particular, when coalescence occurs and when it does not. For a more detailed exposition, the interested reader can check out J Sprittles (2024) \cite{AnnualReview}, where he goes into more detail about the ``missing physics'', and includes the deformation of droplets into consideration, among other things.

Going forward, it would be interesting to see if the equations of motion with the inclusion of the GKE and the vdW forces, as derived in chapters 3.1 and 3.2, have an analytical solution. If not, perhaps we can use numerical methods to approximate when coalescence occurs and when it does not.

\bibliographystyle{plain}
\bibliography{refs.bib}

\end{document}